\documentclass{elsart}
\usepackage{graphics}
\usepackage{epsfig} 
\usepackage{subfigure}
\usepackage{amssymb,latexsym} 
\usepackage{mathrsfs} 
\newcommand{\ea}[1]{\begin{eqnarray}} 
\newcommand{\eb}{\end{eqnarray}}

\def\shiftleft#1{#1\llap{#1\hskip 0.04em}}
\def\shiftdown#1{#1\llap{\lower.04ex\hbox{#1}}} 
\def\thick#1{\shiftdown{\shiftleft{#1}}}
\def\b#1{\thick{\hbox{$#1$}}}
\newcommand{\cal}{\mathscr} 
\begin{document}

\begin{frontmatter} 
\title{Quark model description of quasi-elastic \\ 
pion knockout from the proton at JLAB} 
\author{I.T. Obukhovsky$^1$, D. Fedorov$^1$, Amand Faessler$^2$}, 
\author{Th. Gutsche$^2$ and V.E. Lyubovitskij$^2$} 

\vskip.25cm 
           
\address{$^1$ Institute of Nuclear Physics, 
Moscow State University,\\
Vorobievy Gory, Moscow 119899, Russia\\}  
\address{$^2$ Institut f\"ur Theoretische Physik, 
Universit\"at T\"ubingen, \\ 
Auf der Morgenstelle 14, D-72076 T\"ubingen, Germany}

\date{\today}
 
\maketitle 
 
\vskip.5cm 

\begin{abstract}
The interference term between $s$- and $t$-pole contributions to
the $p(e,e^{\prime}\pi^+)n$ cross section is evaluated on 
the basis of the constituent quark model. It is shown that the
contribution of baryon $s$-poles can be modeled by a nonlocal
extension of the Kroll-Rudermann contact term. This contribution
is in a destructive interference with the pion $t$-pole that is
essential to improve the description of recent JLab data
at the invariant mass $W=$ 1.95 GeV.
Some predictions are made for a new JLab measurement at higher 
values $W=$ 2.1 - 2.3 GeV and $Q^2$ centered at 1.6 and 
2.45~GeV~$^2$/c$^2$.  

\vskip .3cm
 

 
\end{abstract}

\end{frontmatter}

\section{Introduction}

In Ref.~\cite{Volmer:2000ek} the JLab F$\pi$1 Collaboration 
extracted from the data the charged pion form factor $F_{\pi}(Q^2)$ 
using a Regge model for high energy meson 
electroproduction~\cite{Guidal:1997hy}.  Along with this the recent 
JLab data on the Rosenbluth separation of longitudinal and transverse 
cross sections for the reaction $p(e,e^{\prime}\pi^+)n$ were presented
for $Q^2=$ 0.6, 0.75, 1 and 1.6 GeV$^2$/c$^2$ and for the invariant 
mass $W=$ 1.95 GeV.
 
The procedure for reggeization according to~\cite{Guidal:1997hy} is 
quite natural for the meson ($\pi$, $\rho$, $b_1$, ...)
$t$-pole diagrams, however, the authors of Ref.~\cite{Guidal:1997hy} 
have been forced to add a nucleon $s$-pole term to the Regge 
amplitude to ensure gauge invariance. In general, the full sum of 
$s$-channel resonances taken with their proper form factors should 
be dual to the full sum of reggeon exchanges (see, e.g. the recent 
review~\cite{Melnitchouk:2005zr}), but actually the form factors 
for the transitions $\gamma^\ast + N \to N^\ast$ are only known with 
large uncertainties, and thus one can stay on a phenomenological 
level and use only some separate terms from both sums. 

The aim of this letter is to show that an alternative (microscopic) 
description of quasi-elastic pion knockout from the proton can be 
obtained on the basis of a constituent quark model (CQM). Then with
the fixed parameters of model (which are a few in number) one can
predict the absolute value and the $t$-dependence of longitudinal 
$d\sigma_L/dt$ and transverse $d\sigma_T/dt$ cross sections for the 
reaction of pion quasi-elastic knockout from nucleon  by electrons 
in a wide interval of intermediate energies. We start from two 
assumptions, which are natural in the framework of the CQM:

(a) At energies $W\gtrsim$ 2.1 - 2.3 GeV and $Q^2\gtrsim$ 
1 - 2 GeV$^2$/c$^2$ characteristic of the JLab new 
measurements~\cite{Volmer:2000ek,Huber:2003kg} there are no resonance 
peaks in the cross section, and thus one need not to account for the 
details of the intermediate nucleon excitations. Only a combined 
effect of all the $s$- and $u$-channel contributions should be 
evaluated.  In our opinion, a quark approach for such evaluation 
would be more convenient than the traditional baryon-resonance 
approach which characterized by many coupling constants and form 
factors.

The $t$-pole contributions to the quasi-elastic pion knockout 
correspond to trivial quark diagrams shown in Fig.~1a.
Our hypothesis is that in the kinematical region discussed
the full sum of nucleon and excited-nucleon $s$- and $u$-pole terms 
can be represented by the quark diagrams in Figs.~1b,c. 
Moreover, in the region of the pion $t$-pole, $|t|\sim m_{\pi}^2$, 
the sum of all the $s$- and $u$-channel contributions should be 
a small correction to the t-pole contribution 
$\sim (t-m_{\pi}^2)^{-1}.$ Therefore, one can neglect the highest 
order corrections $\sim (m_q^2/Q^2)^n$, e.g. generated by diagrams 
with one- or two-gluon exchange, in comparison to the 
dominant $s$- and $u$-pole terms in Figs.~1b,c.

(b) The $s$- and $u$-pole quark diagrams in Figs.~1b,c at large 
momentum transfer $Q^2$ imply using  the strong and e.-m. quark 
form factors. In the CQM one can take the Fourier transform of the 
pion wave function for the strong $\pi qq$ vertex form factor (e.g. 
this can be shown~\cite{Neudatchin:2004pu} in terms of the $^3P_0$ 
model~\cite{LeYaouanc:1988fx}). Then, the quark amplitude for the 
$s$- or $u$-channel transition $\gamma^\ast+q\to q+\pi$ becomes 
proportional to the product of two form factors, the strong one 
$F_{\pi qq}$ and the electromagnetic form factor of the constituent 
quark $F_q(Q^2)$\footnote{It implies that the constituent quark is 
an extended object and has its own electromagnetic form factor, e.g. 
$F_q(Q^2) = 1/(1+Q^2/\Lambda_q^2)$ ~\cite{Cardarelli:1995hn}. 
The parameter $\Lambda_q$ is believed to be set by the chiral
symmetry scale $\Lambda_{\chi}\simeq 4\pi f_{\pi}\sim$ 1 GeV.}. 
In the considered region of large $Q^2$ and $W$ this product  
is equivalent to the pion electromagnetic form factor 
$F_{\pi}(Q^2)\simeq F_q(Q^2)F_{\pi qq}(Q^2)$. 

As this relationship has not been considered earlier on, here in 
Appendix we obtain it with a simple calculation in terms of two 
diagrams pictured in Fig.~2. The quark diagram for pion form factor 
$F_{\pi}$ is shown in Fig.~2b  with the same notations of momentum 
variables as for the discussed  $s$-pole diagram represented in 
Fig.~2a (with the scalar $\bar qq$ vertex $v$ of $^3P_0$ model). 
The solid line in both pictures shows the propagation of a large 
momentum ${\bf q}\gg{\bf k}$ transfered from photon to pion by 
a highly excited quark. It is intuitively clear that at fixed $q^2$ 
the contribution of this deeply off-shell quark\footnote{In the CQM 
a quark has a fixed mass $m_q\approx M_N/3$ and all the non-trivial
$q^2$-dependence of the quark propagator takes effectively into 
account through the $q^2$-dependent vertex form factors $F_q$ and 
$F_{\pi qq}$.} into the both processes should be similar
in value independently on details in diagrams in Fig.2a and Fig.2b.  
This assumption is formally confirmed by a simple calculation in 
terms of Gaussian wave functions (see Appendix). As a result, in the 
considered kinematical region of quasi-elastic pion knockout both the 
$t$- and $s(u)$-contributions to the amplitude become proportional to 
a common form factor $F_{\pi}(Q^2)$ which can be factorized from the 
sum of $t$- and $s(u)$-pole diagram contributions. Note that in 
a phenomenological approach~\cite{Guidal:1997hy} such common form 
factor was introduced {\it by hand} to preserve gauge invariance 
of the calculated amplitude.

In this work we obtain an evaluation of a common effect of
$s$- and $u$-pole contributions to the cross section. 
The overall contribution of such terms is $Q^2$- and $W$-dependent 
and vanishes with increasing $W$ and $Q^2$ in direct proportion to 
the factor $(QM_N/W^2) \, F_{\pi}(Q^2)$ (the final expression looks 
like a generalization of the Kroll-Ruderman contact term by 
introduction of a $Q^2$- and $W$-dependent form factor into it). 
Our evaluation shows that for JLab data~\cite{Volmer:2000ek} at lower 
$Q^2$ (0.6 and $0.75$ GeV$^2$/c$^2$) this contribution is important 
and cannot be neglected\footnote{Unfortunately at small $Q^2$, 
comparable with the cutoff parameter $\Lambda\simeq$ 0.7 GeV/c in 
$\gamma^\ast NN^\ast$ vertices, one cannot exclude the direct 
contribution of intermediate baryon resonances to the cross section. 
Formally such small $Q^2$ are out of the region of the quasi-elastic 
kinematics.}, while at higher values of $Q^2$  and $W$ it becomes 
considerably smaller and practically invisible in the cross section 
at $W\gtrsim$ 3 GeV. Here we predict the $t$-dependence of the 
longitudinal and transverse cross section for a new JLab measurement 
at $W=$2.1 - 2.3~GeV. 

\section{The $t$- and $s(u)$-channel contributions in terms 
of the CQM}

The possibility of describing the high-energy pion electroproduction  
in terms of the pion $t$-channel mechanism has been discussed in the
literature over many years~\cite{Gutbrod:1973qr,Speth:1995er}. 
An important argument~\cite{Speth:1995er} is that the contribution 
of the $s$-channel diagrams in Fig.1b is suppressed by a factor 
$1/Q^4$ in comparison to the contribution of the $t$-channel 
diagrams. In Refs.~\cite{Speth:1995er,Speth:1996pz,Nikolaev:1994uc}, 
the pion-exchange ($t$-channel diagram) was calculated using the 
light-front dynamics. In principle, this made it possible to extract 
the strong $\pi NN$ form factor $F_{\pi NN}(t)$ from comparison of 
a calculated longitudinal cross section $d\sigma_L/dt$ with the data 
at high momentum transfer $Q^2\gtrsim$ 2 -3 GeV$^2$/c$^2$.
But in practice, the accessible data on high-quality Rosenbluth 
separation are limited by too small $Q^2$ and $W$, where the 
$s(u)$-contributions cannot be neglected.

In contrast to those studies, in Ref.~\cite{Neudatchin:2004pu} 
the process $p(e,e^{\prime}\pi^+)n$ was considered in the laboratory 
frame. In this case one is able to single out, in a natural way, 
the kinematical region where the recoil momentum of the final nucleon 
$P^{\prime\mu}=\{P_0^{\prime},-{\bf k}\}$ with 
$P_0^{\prime}\simeq M_N+{\bf k}^2/(2M_N)$ is small and where only the 
momentum ${\bf k}^{\prime}={\bf q}+{\bf k}$ of the knocked out meson 
is large. In that (quasi-elastic) region at 
$|t|\lesssim 0.1 - 0.2$~GeV$^2$ 4-momentum $k=P-P^{\prime}$ 
transfered to the nucleon can be related to the 3-momentum ${\bf k}$, 
$k^{\mu}\simeq \{-{\bf k}^2/(2M_N), {\bf k}\}$.  
Therefore, both the initial $|N(P)>$ and the final-state
$|N(P^{\prime})>$ nucleons are nonrelativistic ones. In the CQM 
each of them can be described by a nonrelativistic wave function 
\begin{eqnarray}
|N_{3q}(P) \rangle &=&\Phi_N({\b\rho}_1,{\b\rho}_2)\,
|[2^3]_C[3]_{SI}S_z,I_z \rangle\,. 
\label{nuc}
\end{eqnarray}
Here, $|[2^3]_C[3]_{SI}S_z,I_z>$ is the 
color$(C)$-spin$(S)$-isospin$(I)$ part and 
\begin{eqnarray}
\Phi_N({\b\rho}_1,{\b\rho}_2))=
{\cal N}^{-1}_N e^{-{\b\rho}^2_1/4b^2-{\b\rho}_2^2/3b^2} 
\label{nuc1} 
\end{eqnarray} 
is the internal wave function normalized as 
$\int d{\b\rho}_1d{\b\rho}_2
\vert\Phi_N({\b\rho}_1,{\b\rho}_2)\vert^2=1$,   
where ${\b \rho}_1$ and ${\b \rho}_2$ are the Jacobi coordinates, 
${\cal N}_N$ is a normalization constant and $b$ is 
a ``quark radius'' of the nucleon. 

The $t$-pole amplitude for the $\gamma^\ast q\to q^{\prime}\pi^+$ 
process on the $i$-th quark (Fig.1a) is 
\begin{eqnarray} 
{\cal M}^{(i)\mu}_{q(t)} \, = \, ie \, \hat G^{(i)}_{\pi qq}
\, \frac{F_{\pi}(Q^2)}{t-m_{\pi}^2} \, (k+k^{\prime})^\mu \,, 
\hspace*{.5cm} 
\hat G^{(i)}_{\pi qq} = g_{\pi qq}\tau_{-}^{(i)}{\b\sigma}^{(i)} 
\cdot{\bf k}, 
\label{tp}
\end{eqnarray} 
where $F_{\pi}(Q^2)$ is the pion electromagnetic form factor and
$\hat G^{(i)}_{\pi qq}$ is the $\pi qq$ vertex operator for 
the $i$-th quark, which is related to the pion-nucleon form factor 
$G_{\pi\scriptscriptstyle{ NN}}({\bf k}^2)$ as  
\begin{eqnarray}
\hspace*{-.5cm}
G_{\pi\scriptscriptstyle{ NN}}({\bf k}^2)\tau_-{\b\sigma}\cdot{\bf k} 
\equiv\langle N_{3q}(P^{\prime})|\sum\limits_{i=1}^{3} 
\hat G^{(i)}_{\pi qq}|N_{3q}(P) \rangle 
=\frac{5\tau_-}{3} \, g_{\pi qq}\, {\b\sigma}\cdot{\bf k} \, 
e^{-{\bf k}^2b^2/6}. 
\label{cqm}
\end {eqnarray}
In the following, for convenience, 
we proceed with the normalized pion-nucleon form factor 
$F_{\pi\scriptscriptstyle{ NN}}({\bf k}^2) \doteq 
G_{\pi\scriptscriptstyle{ NN}}({\bf k}^2)/
g_{\pi\scriptscriptstyle{ NN}}$ 
with $g_{\pi\scriptscriptstyle{ NN}} \equiv 
G_{\pi\scriptscriptstyle{ NN}}(0)$.  
We use a simple form of the operator $\hat G^{(i)}_{\pi qq}$ 
neglecting the momentum dependence since the exchanged pion 
and the constituent quarks are approximately on their mass shells.

The quark-level amplitude ${\cal M}^{\mu}_{q(t)}$ gives rise to
the $t$-pole matrix element for $\pi^+$ electroproduction from 
the nucleon
\begin{eqnarray}
{\cal M}^{\mu}_{N(t)}=\langle N_{3q}(P^{\prime})|
\sum\limits_{i=1}^{3} {\cal M}^{(i)\mu}_{q(t)}
|N_{3q}(P) \rangle\,, 
\label{tpme}
\end{eqnarray}  
which is proportional to the strong form factor 
\begin{eqnarray}
F_{\pi NN}(t)=
\int d{\b\rho}_1d{\b\rho}_2e^{i\frac{2}{3}{\bf k}\cdot{\b\rho}_2}
\vert\Phi_N({\b\rho}_1,{\b\rho}_2)\vert^2=e^{-{\bf k}^2b^2/6}\,,
\quad t\simeq-{\bf k}^2\,,
\label{fpnn}
\end{eqnarray}  
defined above.  
At $b = 0.6$ fm, which is a typical scale in the CQM (at small 
$|t| \lesssim 0.2$ GeV$^2$ this Gaussian is very close to
the monopole form factor with $\Lambda_{\pi NN}\simeq 0.7$ GeV), 
the matrix element (\ref{tpme}) gives a reasonable description of the
$t$-dependence of the differential cross sections in the 
JLab experiment (see, e.g., Ref.~\cite{Neudatchin:2004pu}). 
However, since the $s$- and $u$- channel quark contributions 
(Fig.~1b,c), which are required for gauge invariance, have not been 
taken into account, the results of such  simple quark evaluation 
(dashed lines in the upper panels of Fig.~3) are not so good in 
comparison with the Regge parametrization~\cite{Guidal:1997hy} 
used in Refs.~\cite{Volmer:2000ek,Volmer:2000vl} (dash-dotted lines 
in Fig.~3). 

Here we shall evaluate these contributions, starting from 
a nonlocal (nl) variant of the pseudoscalar $\pi qq$ coupling 
$g_{\pi qq}^{\rm nl}$, 
which differs from the local coupling $g_{\pi qq}$ by the presence 
of the quark form factor: 
$g_{\pi qq}^{\rm nl}(p^2)  \doteq g_{\pi qq} F_{\pi qq}(p^2)$.   
Hence, the vertex operator $\hat G^{(i)}_{\pi qq}$ is modified 
accordingly. In this case the kinematics is the following.  
Two particles, pion and ingoing (outgoing) quark, are on mass shell:  
${k^{\prime}}^2=m_{\pi}^2$, $p_i^2=m_q^2$ (${p_i^{\prime}}^2=m_q^2$), 
where $m_q=M_N/3$ is the mass of the constituent quark.
But the intermediate quark with a large momentum
$p_{i(s)}^{\prime\prime}=p_i+q$ 
($p_{i(u)}^{\prime\prime}=p^{\prime}_i-q$)
obtained by absorption (emission) of a momentum transfer from the 
$\gamma^\ast$ should be considerably off its mass shell: 
${p_{i}^{\prime\prime}}^2\simeq Q^2$, (see solid lines in Figs.1b,c).

The contribution of the quark diagrams in Figs.1b,c to the matrix
element for absorption of the longitudinal virtual photon
($\epsilon^{(0)\,\mu}=
\frac{1}{Q}\{|{\bf q}|,\, q_0\hat{\bf q}\}$, $Q=\sqrt{-q^2}$) 
\begin{eqnarray}
{\cal M}^{(i)\mu}_{q(s + u)} &=& - \, i \, \sqrt{2}\, g_{\pi qq} \, 
F_{\pi qq}(Q^2) \, F_q(Q^2) \, 
\nonumber \\ 
&\times&\left\{ e_u 
\gamma^5\frac{1}{
\not\! p^{\,\,\prime}_i+\! \not k^{\prime} 
-m_q}\gamma^{\mu} 
\, + \, e_d  \gamma^{\mu}\frac{1}{\not\!p_i-\!
\not k^{\prime}-m_q}\gamma^5 \right\}^{(i)}
\label{sp} 
\end{eqnarray}
was evaluated in the approximation 
$W^2 \gg (Q^2-M_N^2),\,M_N^2$. Here 
$e_u(e_d)$ is the quark charge and $F_q(Q^2)$ is 
the quark electromagnetic form factor. 
In the considered region of a large momentum 
$k^{\prime \mu} = (k^{\prime 0}, {\bf k^\prime})$ 
transfered to the pion (with 
$k^{\prime 0}\simeq |{\bf k}^{\prime}|\simeq|{\bf q}|$) 
and a small momentum 
$k^\mu=\{t/(2M_N),{\bf k}\}\simeq\{0,{\bf k}\}$ 
transfered to the nucleon we can omit all contributions 
to the amplitude proportional to small parameters
\begin{eqnarray}
\frac{Q^2-M_N^2}{W^2},\,\,\frac{|{\bf k}|}{2M_N},\,\,
\frac{m_q^2}{W^2},\,\,\frac{m_{\pi}^2}{M_N^2},\,\dots\ll 1\,. 
\label{sml}
\end{eqnarray}
Then, for the 4-momenta of intermediate (off-shell) quarks in 
the diagrams in Fig.~1b,c one can approximately write
\begin{eqnarray}
p_{i(s)}^{\prime\prime\,2}
\simeq 2m_qk_0^{\prime}\left(
1+\frac{|{\b\varkappa}_2|}{m_q}cos\theta\right)-Q^2,\quad
&p_{i(u)}^{\prime\prime\,2}\simeq  -p_{i(s)}^{\prime\prime\,2}-Q^2,
\label{p3su}
\end{eqnarray}
where ${\b\varkappa}_2$ is a relative 
momentum conjugated to the Jacobi coordinate ${\b\rho}_2$, and
$cos\theta=\hat{\bf k}^{\prime}\cdot\hat{\b\varkappa}_2
\simeq \hat{\bf q}\cdot\hat{\b\varkappa}_2$. 
By making use of the approximate equality
$|{\bf q}|/q_0 \simeq 1 + Q^2/(2q^2_0)$, which holds within 
the conditions~(\ref{sml}), one obtains the approximation
$({\not k}^{\prime}\gamma^5\gamma^{\mu}
+\gamma^5\gamma^{\mu}{\not k}^{\prime})\varepsilon^{(0)}_{\mu}\simeq
-k_0^{\prime}Q/q_0\,{\rm diag}\{{\b\sigma}\cdot\hat{\bf q},
{\b\sigma}\cdot\hat{\bf q}\}$ useful for reducing
the r.h.s. of Eq.~(\ref{sp}) (the approximation implies that 
$\hat {\bf k}^{\prime}\simeq \hat{\bf q}$, 
which is true for the forward pion knockout).

Finally, in the lowest order of expansion in the small parameters 
$M_N^2/W^2$ and $Q^2/W^2$ the matrix element~(\ref{sp}) reduces to an
effective $\gamma\pi qq$ interaction\footnote{At the photon 
point $Q^2=$0, $\epsilon^{(\lambda)}_{\mu}=
\{0,\,\hat{\bf n}_{\perp}\}$ and in the low-energy limit 
$k^{\prime}_0\to m_{\pi}$
the same quark calculation gives exactly rise to the Kroll-Ruderman 
term~\cite{Kroll:1953vq} for the threshold pion photoproduction.}
\begin{eqnarray}
{\cal M}_{q(s+u)}^{(i)\mu}\epsilon^{(\lambda\!=\!0)}_{\mu}&=&
-i\frac{eg_{\pi qq}}{2m_q}\,
\frac{M_NQ}{W^2}\,
\,\frac{\tau^{(i)}_-({\b\sigma}^{(i)}\cdot\hat{\bf q})}
{(1+\frac{|{\b\varkappa}_2|}{m_q}cos\theta)}\, 
F_q(Q^2)F_{\pi qq}(Q^2), 
\label{me}
\end{eqnarray}
in which one can use the relation
\begin{eqnarray}
F_q(Q^2)F_{\pi qq}(Q^2)\simeq F_{\pi}(Q^2), 
\label{fqq}
\end{eqnarray}
discussed in the Section 1 (see also Appendix).
Note that the ${\b\varkappa}_2$-dependent denominator in 
Eq.~(\ref{me}), being integrated 
with Gaussian functions (\ref{nuc}), introduces only a
small correction to the integral, and thus we shall omit it
in the final expression (see next section) for the nucleon 
matrix element (but it has not been omitted in the numerical
calculation). 

Eq.~(\ref{fqq}) is very useful to check gauge invariance of
the sum of $t$- and $s(u)$-pole amplitudes in the final expression,
where these amplitudes destructively interfere (see Sect. 3).
It should be noted that gauge invariance can also be 
restored in the general case, if one uses different vertex 
form factors (see detailed discussion in 
Refs.~\cite{Koch:2001ii,Gross:1987bu}).  

\section{Longitudinal and transeverse cross section}

We calculate the longitudinal part of differential cross section  
\begin{eqnarray}
\frac{d\sigma_L}{dt}=
\frac{M_N^2\overline{|J_{\pi}^{\mu}
\varepsilon^{(\lambda=0)}_{\mu}|^2}}
{4\pi(W^2-M_N^2)\sqrt{(W^2-Q^2-M_N^2)^2+4W^2Q^2}}
\label{sl}
\end{eqnarray}
for quasi-elastic pion knockout taking into account both the $t$-pole 
diagram in Fig.1a and the $s$- and $u$-pole contribution~(\ref{me}), 
\begin{eqnarray} 
\overline{|J_{\pi}^{\mu}\varepsilon^{(\lambda=0)}_{\mu}|^2}=
\frac{1}{2}\sum_{spin}\vert\langle N_{3q}(P^\prime)|\,\sum_{j\!=\!1}^3
[{\cal M}^{(j)\mu}_{q(t)}+{\cal M}^{(j)\mu}_{q(s + u)}]
\varepsilon^{(\lambda=0)}_{\mu}|N_{3q}(P) \rangle\vert^2\,, 
\label{je}
\end{eqnarray}
and compare the results to the JLab data~\cite{Volmer:2000ek} and to 
the Regge model~\cite{Guidal:1997hy} 
predictions~\cite{Volmer:2000ek,Volmer:2000vl} (see Fig.~3).

The CQM calculation of the right-hand side of Eq.~(\ref{je})
leads to a simple matrix element for a nonrelativistic 
nucleon wave function $\Phi_N({\b\rho}_1,{\b\rho}_2)$. 
Using Eqs.~(\ref{tpme}) and (\ref{me}) we  obtain in the laboratory 
frame the following matrix element of the longitudinal 
hadron current for the transition $\gamma^\ast+p\to n+\pi^+$: 
\begin{eqnarray}
J_{\pi}^{\mu}\varepsilon^{(0)}_{\mu}=i\tau_{-}
\frac{eg_{\pi{\scriptscriptstyle NN}}}{2M_N}F_{\pi}(Q^2)
F_{\pi NN}(t)\biggl[
\frac{2 \varepsilon^{(0)} \cdot k^\prime \b\sigma 
\cdot {\bf k}}{t-m_{\pi}^2}
-\frac{M_NQ}{W^2}{\b\sigma}\cdot\hat{\bf q}\biggr]\,.
\label{je1}
\end{eqnarray}
Here the contribution of the $b_1$-meson pole (the P-wave 
excitation of the pion in the CQM) may be also taken into account. 
Then, the third term,
$+(g_{b_1{\scriptscriptstyle NN}}/g_{\pi{\scriptscriptstyle NN}})\,
(g_{b_1\pi\gamma}k_0Q/(2M_{\scriptscriptstyle N}m_{b_1}))\,
({\bf k}\cdot\hat{\bf q}\,{\b\sigma}\cdot{\bf k}/(t-m_{b_1}^2))$,
should be inserted into the square brackets in the r.h.s. 
of Eq.~(\ref{je1}). However, the $b_1$-pole contribution 
is too small when compared to the $\pi$-pole contribution. 
The $\rho$-meson pole is also not included as in the CQM 
it does not contribute to the longitudinal part of the cross section 
(in the CQM the quark spin-flip amplitude 
$\rho^++\gamma^\ast(M1)\to\pi^+$ 
only contributes to the transverse part of the cross section where it 
is of prime importance~\cite{Speth:1995er,Obukhovsky:2005ob}).
 
The quark model calculation leading to the r.h.s. of Eq.~(\ref{je1}) 
has the advantage that
the parameters of the vertices $g_{\pi NN}$,  $g_{b_1 NN}$, 
$g_{b_1\pi\gamma}$ and their form factors $F_{\pi NN}(t)$, 
$F_{b_1 NN}(t)$, $F_{b_1}(Q^2)$ are not free, but related to 
each other by the following relationships:
(i) The integral part of Eq.~(\ref{je}) defines a nonrelativistic 
vertex form factor~(\ref{fpnn}) 
$F_{m NN}(t)=F_{\pi NN}(t)=F_{b_1 NN}(t)$ 
common to all the terms in Eq.~(\ref{je1}).  
(ii) The relative phases of all the amplitudes in the r.h.s of 
Eq.~(\ref{je1}) are fixed by the results of quark model calculations. 
Therefore, the negative sign of the interference term between 
$t$- and $s$-pole contributions (the destructive interference) 
is unambiguously determined: the spin average of the product 
of the first term in the squared bracket of 
Eq.~(\ref{je1}) with the second one has a negative value: 
$1/2 \, \sum_{S_zS_z^{\prime}}
<S_z^{\prime}|{\b\sigma}\cdot\hat{\bf k}|S_z>
<S_z|{\b\sigma}\cdot\hat{\bf q}|S_z^{\prime}>=
\cos(\widehat{{\bf k}{\bf q}})\,.$
Recall that $\cos(\widehat{{\bf k}{\bf q}}) \simeq \,$-1 in 
the quasielastic pion knockout kinematics, where $-{\bf k}$ 
is the momentum of the nucleon-spectator in the lab frame.  
 
The $\rho$-pole contribution to the transverse part of differential 
cross section $d\sigma_T/dt$ is proportional to the value 
$\overline{|J_{\rho}^{\mu}\varepsilon^{(\lambda=1)}_{\mu}|^2}$, where
\begin{eqnarray}
J_{\rho}^{\mu}\varepsilon^{(\lambda)}_{\mu}=
i\tau_-eg_{\rho\pi\gamma}\frac{M_{\rho}}{M_{\pi}}
F_{\rho\pi\gamma}(Q^2)\frac{1+\varkappa_{\rho}}{2M_N}
g_{\rho{\scriptscriptstyle NN}}F_{\rho{\scriptscriptstyle NN}}(t)
|{\bf q}|\frac{i[{\b\sigma}\times{\bf k}]
\cdot{\b\epsilon}^{(\lambda)}}{t-M_{\rho}^2}
\label{je2}
\end{eqnarray}
and $\varepsilon^{(\lambda)\mu}=\{0,{\b\epsilon}^{(\lambda)}\}$
for $\lambda=\pm$ 1. For coupling constants and form factors
in Eq.~(\ref{je2}) we use the results of calculations
~\cite{Obukhovsky:2005ob} in terms of $^3P_0$ model from
Ref.~\cite{Neudatchin:2004pu}:
$g_{\rho{\scriptscriptstyle NN}}=\frac{1}{5}
g_{\pi{\scriptscriptstyle NN}}\sqrt{\frac{M_{\rho}}{M_{\pi}}}$, 
$1\!+\!\varkappa_{\rho}=$ 5, 
$F_{\rho{\scriptscriptstyle NN}}(t)=
F_{\pi{\scriptscriptstyle NN}}(t)$. 
For $g_{\rho\pi\gamma}$ coupling constant we take the 
value $g_{\rho\pi\gamma} = 0.103$ which is fixed 
by the the experimental value of the 
decay width $\Gamma_{\rho\to\pi\gamma}=$ 67 KeV. The momentum 
dependence of $F_{\rho\pi\gamma}(Q^2)$ will be discussed in the 
next section. 

\section{Results and outlook}

In our calculation we use a standard (monopole-like) 
representation for the transition form factors $F_{\pi}(Q^2)$ 
and $F_{\rho\pi\gamma}(Q^2)$: 
\begin{eqnarray} 
F_{\pi}(Q^2) = \frac{1}{1 + Q^2/\Lambda^2_{\pi}}\,, 
\quad\quad\quad
F_{\rho\pi\gamma}(Q^2) = \frac{1}{1 + Q^2/\Lambda^2_{\rho\pi}}\,.  
\end{eqnarray}  
For the pion charge form factor 
we use $\Lambda^2_{\pi}=$0.54 GeV$^2$/c$^2$ which is close to 
the recent theoretical evalu\-a\-ti\-on~\cite{Maris:2000sk} 
and correlates well with the recent JLab data~\cite{Volmer:2000ek}. 
In the case of the $F_{\rho\pi\gamma}(Q^2)$ form factor we 
vary $\Lambda^2_{\rho\pi}$ from $\Lambda^2_{\pi}=0.54$~GeV$^2$/c$^2$ 
to 0.7 GeV$^2$/c$^2$. We think that an accurate analysis of the 
transverse part of the differential cross section can shed light 
on the value of $\Lambda_{\rho\pi}$. From Fig.~3 one can see, 
that the value $\Lambda^2_{\rho\pi}=0.7$ GeV$^2$/c$^2$ is more 
appropriate in description of the transverse cross section.
Apart from this value we vary only one free parameter in 
the standard representation of the strong $\pi NN$ form factor
$F_{\pi NN}(Q^2)=\Lambda_{\pi NN}^2/(\Lambda^2_{\pi NN}\!+\!Q^2)$ 
where the range parameter $\Lambda_{\pi NN}\simeq$ 0.6 - 0.7 GeV/c 
corresponds to the reasonable value  $b\simeq$ 0.5 - 0.6 fm for 
the radius of the three-quark configuration $s^3$ 
used in the CQM nucleon wave function~(\ref{nuc}).
At realistic values of $\Lambda_{\pi NN}=0.7$~GeV/c and 
$g_{\pi NN}=13.5$ our results (the solid lines in Fig.3) are 
in a better agreement with the data~\cite{Volmer:2000ek} than 
the simplified model~\cite{Neudatchin:2004pu} (dashed lines in two
upper panels) taking into account only the pion $t$-pole not 
satisfying gauge invariance.

In our calculation we model the contribution of baryon resonances 
in the $s$- and $u$-channel by a nonlocal extension of the 
Kroll-Rudermann contact term. This contribution is negative, which is 
essential to improve the description of the measured cross section.  
At intermediate values of $Q^2 \gtrsim$ 1 GeV$^2$/c$^2$, 
which correspond to a quasi-elastic mechanism of pion knockout, 
the calculated cross section is close to the experimental data. 
At smaller $Q^2\lesssim$ 0.7  GeV$^2$/c$^2$ our results are close 
to the cross section  calculated 
in Ref~\cite{Volmer:2000ek,Volmer:2000vl} (dash-dotted lines) 
on the basis of a model~\cite{Guidal:1997hy} which takes into account 
both the Reggeon-pole exchange and the nucleon-pole  
contribution.  

For a new JLab measurement of $d\sigma_L/dt$ and $d\sigma_T/dt$ at 
higher invariant mass (the data analysis is presently underway) in 
Fig.~4 we give our prediction for the $t$-dependence of the cross 
sections at $W=$ 2.1 - 2.3 GeV and $Q^2$ centered at 1.6 and 2.45  
GeV$^2$/c$^2$. Here we use fixed (above defined) parameters of the 
model. Because of the large value of $W$ the  contribution of the 
effective contact term~(\ref{me}), which is proportional to the 
factor $QM_N/W^2$, becomes too small and not shown in Fig.~4.   

Our calculations show that the longitudinal cross section practically
does not depend on the contribution of the $b_1$-meson pole, and 
thus the data~\cite{Volmer:2000ek} cannot constrain the $b_1NN$ and 
$b_1\pi\gamma$ vertices. In contrast, data on the transverse cross
section should be critically dependent on the $\rho$-pole contribution
and on the $\rho\pi\gamma^\ast$ spin-flip amplitude. This was first 
shown in Ref.~\cite{Speth:1995er} and then supported in 
Ref.~\cite{Obukhovsky:2005ob}. Our results (Figs.~3 and 4) 
confirm this statement and show that the data on the $d\sigma_T/dt$
at high $Q^2$ and $W$ can be used for the direct measurement
of the $F_{\rho\pi\gamma}(Q^2)$ form factor. 
On the other hand, a Regge description~\cite{Guidal:1997hy} of the 
transverse cross section is not as good as for the longitudinal case. 

\newpage

{\bf Acknowledgements}

\noindent 
The authors thank the Fpi2 Collaboration 
[Experiment E01-004 at JLab] for the interest to 
our work and informative discussion. 
We thank Vladimir Neudatchin, Nikolai Yudin,  
Garth Huber and Tanja Horn for fruitful discussions 
and suggestions.
This work was supported by the DFG under con\-tra\-cts FA67/25-3 
and GRK683 and the DFG grant 436 RUS 113/790/. 
This research is also part of the EU Integrated 
Infrastructure Initiative Had\-ron\-phy\-si\-cs project under 
contract number RII3-CT-2004-506078, 
President grant of Russia "Scientific Schools"  No. 1743.2003 and 
grants of Russia RFBR No. 05-02-04000, 05-02-17394 and 03-02-17394.

\appendix

\section{Appendix}
The contribution of the diagram in Fig.~2a is proportional 
to the overlap integral~\cite{Neudatchin:2004pu}
\begin{eqnarray}
F_{diag}({\bf q},{\bf k})&=&F_q({\bf q}^2)\frac{1}{\cal N}\!
\int\frac{d{\b\varkappa}_2}{(2\pi)^3}
\Phi_N({\b\varkappa}_2)\Phi_N\biggl({\b\varkappa}_2\!
+\!\frac{2}{3}{\bf k}\biggr)
\Phi_{\pi}\biggl(\!-\!{\b\varkappa}_2\!
+\!\frac{{\bf q}\!-\!{\bf k}}{2}\biggr)
\nonumber\\
&=&e^{-{\bf k}^2b^2/6}F_q({\bf q}^2) \, 
{\rm exp}\left[-\frac{({\bf q}\!-\!\frac{\bf k}{3})^2
b_{\pi}^2}{4(1+2x_{\pi}^2/3)}
\right],
\label{fqk}
\end{eqnarray}
where $\Phi_N$ is a wave function of the 3-rd quark in the nucleon
and $\Phi_{\pi}$ is a pion wave function in the momentum 
representation (here it plays the role of a strong $\pi qq$ 
form factor). In this simple calculation we use 
$\Phi_N({\b\varkappa}_2)\sim e^{-3{\b\varkappa}_2^2b^2/4}$ 
and $\Phi_{\pi}({\b\varkappa})\sim e^{-{\b\varkappa}^2b_{\pi}^2}$,  
where 
${\b\varkappa}_2=({\bf p}_1\!+\!{\bf p}_2\!-\!2{\bf p}_3)/3$. 
Here $b$ and $b_{\pi}$ are the quark radii in nucleon and pion 
correspondingly with $x_{\pi}=b_{\pi}/b$. In the CQM one can obtain 
a very similar expression for the diagram in Fig.~2b:
\begin{eqnarray}
\hspace*{-.5cm}F_{\pi}({\bf q}^2)&=&F_q({\bf q}^2)
\int\frac{d{\bf p}}{(2\pi)^3}\Phi_{\pi}\biggl({\bf p}\!
-\!\frac{{\bf k}}{2}\biggr)
\Phi_{\pi}\biggl({\bf p}\!+\!\frac{{\bf q}\!-\!{\bf k}}{2}\biggr)
= F_q({\bf q}^2)e^{-{\bf q}^2b_{\pi}^2/8} \, .
\label{fqpi}
\end{eqnarray} 
One can see that the strong $\pi NN$ form factor (in the CQM it is  
$F_{\pi{\scriptscriptstyle NN}}({\bf k}^2)=e^{-{\bf k}^2b^2/6}$) 
appears as a multiplier in the second line of 
Eq.~(\ref{fqk}) and the reminder of this expression coincides with 
the pion form factor (\ref{fqpi}) in the limit ${\bf q}\gg{\bf k}$
at a specific value $x_{\pi}=\sqrt{3/2}$, which is not very 
different from the realistic value $x_{\pi}\approx$ 1 characteristic 
of the CQM. Really the factor $\sqrt{3/2}$ appears because
of a specific procedure of center-of-mass motion eliminating, which
is model dependent. We shell consider  a small difference between
$F_{\pi}$ and ${F_qF_{\pi qq}}$ as a small model-dependent correction
which can be neglected in line of the assumption (a) given in 
Section 1.

It could also happen that the $\gamma^*$ is absorbed by one quark 
while the pion is emitted from a different quark in the nucleon. 
Then a large momentum ${\bf q}$ should be 
firstly transfered to the $N$-$3q$ vertex. 
In the CQM this amplitude can be evaluated as well. 
One can insert an intermediate nucleon (baryon) state between two
points in the diagram where a large momentum 
${\bf q}$ or  $|{\bf k}^{\prime}|\approx|{\bf q}|$ is absorbed or
emitted. It is essential that the nucleon 
propagator and form factors (both electromagnetic and strong)
depend on the large momentum. This leads to the following contribution 
to the amplitude of pion quasi-elasic knockout, which scales as: 
\begin{eqnarray}
F_{nondiag}({\bf q},{\bf k}^{\prime})\sim F_q({\bf q}^2)
\int\frac{d{\b\varkappa}_2}{(2\pi)^3}\Phi_N({\b\varkappa}_2)
\Phi_N\biggl({\b\varkappa}_2\!+\!\frac{2}{3}{\bf q}\biggr)
\frac{M_N \, \sqrt{Q^2}}{Q^2-M_N^2}\nonumber\\
\times\int\frac{d{\b\varkappa}^{\prime}_2}{(2\pi)^3}
\Phi_N({\b\varkappa}^{\prime}_2)
\Phi_N\biggl({\b\varkappa}^{\prime}_2\!-\!
\frac{2}{3}{\bf k}^{\prime}\biggr) \approx G_E(Q^2)
\frac{M_N}{\sqrt{Q^2}} F_{\pi NN}(Q^2)\,,
\label{fnd}
\end{eqnarray}
where $G_E$ is a nucleon electric form factor.
At large $Q^2\gtrsim$ 1 - 2 GeV$^2$/c$^2$ characteristic of the new
F$\pi$2 measurement the ratio 
$|F_{nondiag}({\bf q},{\bf k}^{\prime})/F_{diag}({\bf q},{\bf k})|^2$ 
is too small. Hence one 
can consider the amplitude (\ref{fnd}) as a second order correction
to the $\pi$-pole amplitude, while the amplitude (\ref{fqk}) is the
first order correction.

\vspace*{.5cm}

\centerline{LIST OF FIGURES}

\vspace*{.1cm}

\noindent Fig.1: Quark diagrams for the $t$-, $s$- and $u$-channel 
mechanism of pion knockout. Thick lines indicate large $Q^2$ 
transition. 

\vspace*{.15cm}

Fig.2: Quark diagrams: (a) for the $s$-pole mechanism of pion
emission (in terms of $^3P_0$ model), (b) for pion form factor.

\vspace*{.15cm}

Fig.3: Longitudinal and transverse
cross sections for the $p(e,e'\pi^+)n$ process.
The $F\pi1$ data~\cite{Volmer:2000ek} for $W\simeq$ 0.95 GeV. 
1) The upper row of panels: $Q^2=$ 0.6 GeV$^2$/c$^2$ (left 
panel) and 0.75 GeV$^2$/c$^2$ (right panel). 
Dashed lines: the pion $t$-pole only. 
Solid lines: the total sum of the pion $t$-pole 
and the contact ($s+u$)-term with common e.-m. and strong
form factors.  Dotted lines: the same sum, but
without form factors in the contact term. 
Dash-dotted lines: the Regge model~\cite{Guidal:1997hy} 
prediction. 
2) The lower two rows of panels: $Q^2=$ 1.0 GeV$^2$/c$^2$
(upper panels) and 1.6 GeV$^2$/c$^2$ (lower panels). 
Left panels ($d\sigma_L/dt$): 
the same notations as in the first two panels. 
Right panels ($d\sigma_T/dt$): the $\pi$-pole contribution 
only (dotted); the sum of $\pi$- and $\rho$-pole + ($s+u$) 
contributions with a monopole $\rho\pi\gamma$ form factor, 
$\Lambda_{\rho\pi}^2=$ 0.54 GeV$^2$/c$^2$ (solid) and 0.7 
GeV$^2$/c$^2$ (dashed).

\vspace*{.15cm}

Fig.4: Predicted cross sections for new JLab 
measurements ~\cite{Huber:2003kg} at a higher value 
of $W=$ 2.1 - 2.3 GeV and  $Q^2$ centered at 1.6 GeV$^2$/c$^2$
(upper curves) and 2.45 GeV$^2$/c$^2$ (lower curves). 
Left panel ($d\sigma_L/dt$): the total sum of the pion 
$t$-pole and the contact ($s+u$)-term (solid); the Regge
model prediction (dash-dotted). Right panel ($d\sigma_T/dt$): 
the $\pi$-pole contribution only (dotted);
the sum of $\pi$- and $\rho$-pole + ($s+u$) contributions with a 
monopole $\rho\pi\gamma$ form factor, $\Lambda_{\rho\pi}^2=$ 0.54 
GeV$^2$/c$^2$ (solid) and 0.7 GeV$^2$/c$^2$ (dashed). 

\newpage

\begin{figure}

\vspace*{1.5cm} 

\centering{\
\epsfig{figure=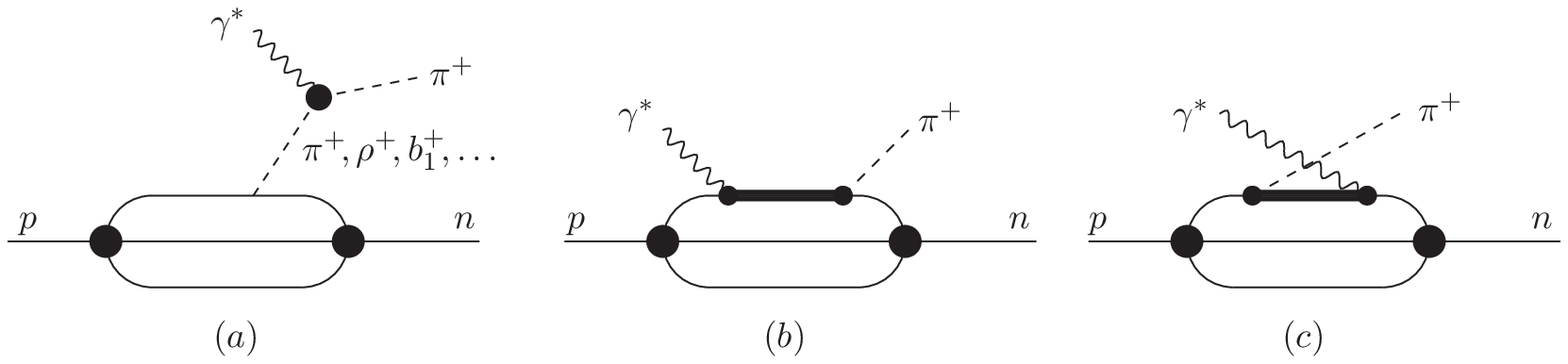,scale=.85}} 

\vspace*{-5.5cm}

\centerline{\bf Fig.1}

\centering{\
\epsfig{figure=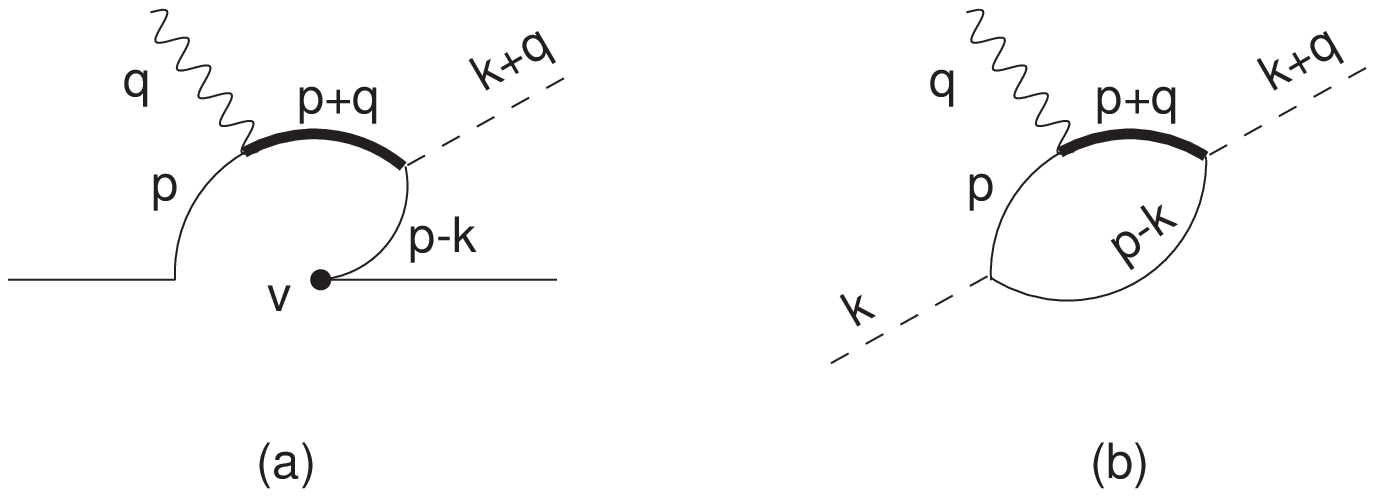,scale=.65}}

\vspace*{-1cm}

\centerline{\bf Fig.2}

\end{figure}

\newpage

\begin{figure}

\vspace*{-1cm}
\centering{\
\epsfig{file=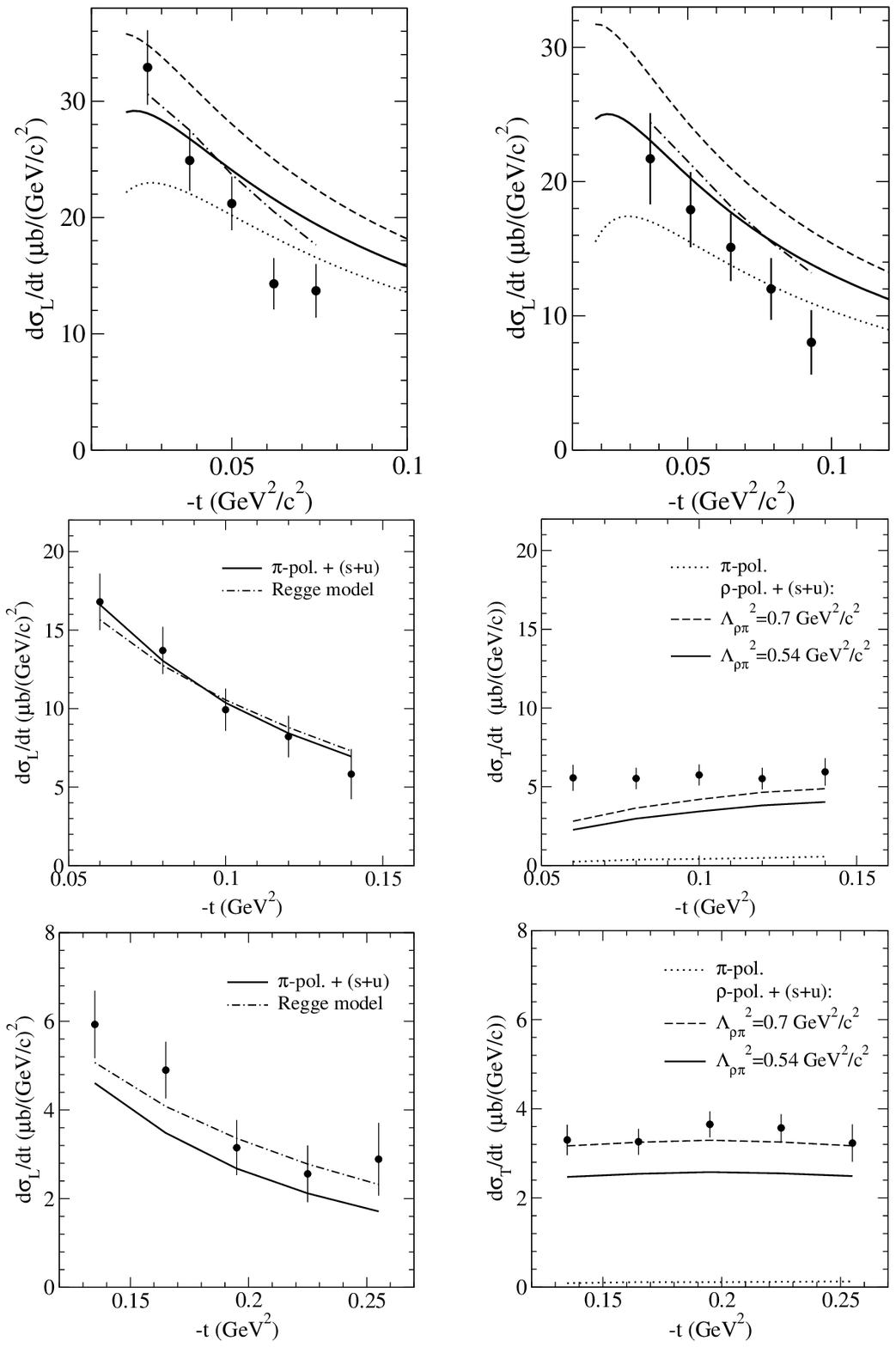,scale=0.65}}

\vspace*{-4.5cm}

\centerline{\bf Fig.3}
\end{figure}

\begin{figure}
\centering{\
\epsfig{file=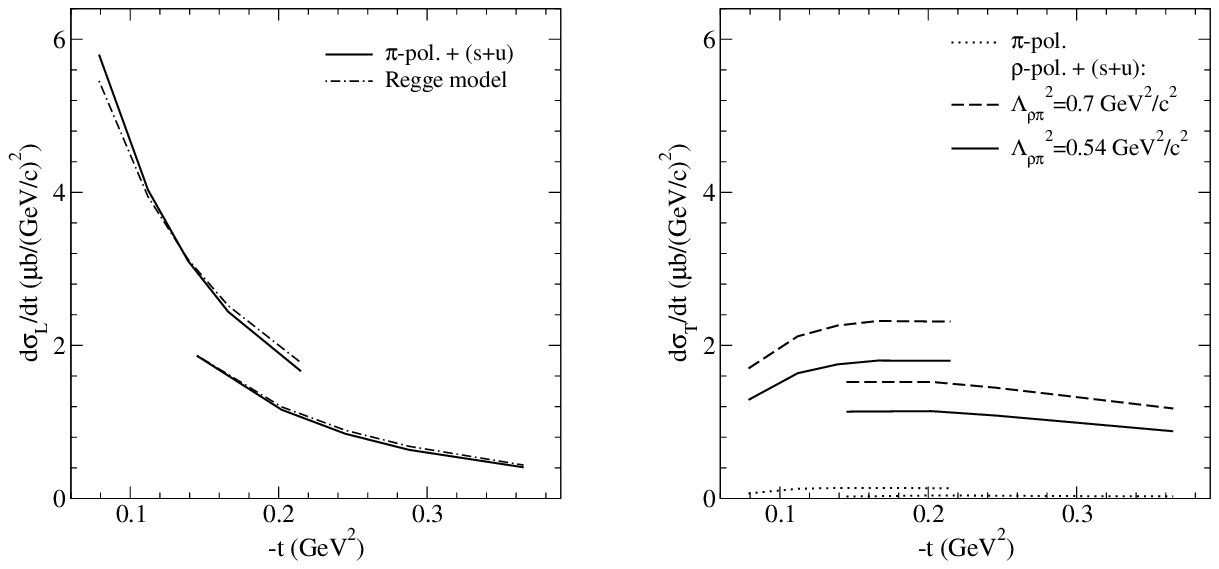,scale=0.65}}

\vspace*{-4cm}

\centerline{\bf Fig.4}
\end{figure}

\end{document}